\documentclass[a4paper,12pt]{article}
\pdfoutput=1
\usepackage{graphicx,subfigure,amsmath,amssymb}
\usepackage{cite}
\usepackage{makecell}
\allowdisplaybreaks[4]

\newlength{\dinwidth}
\newlength{\dinmargin}
\usepackage{color}

\begin{document}

\title{\bf  Study of  $\bar{B}_{u,d,s}^* \to D_{u,d,s}^* V\,(V=D_{d,s}^{*-}\,,K^{*-}\,,{\rho}^-)$  weak decays }
\author{Qin Chang$^{a,b}$, Yunyun Zhang$^{a}$ and Xiaonan Li$^{a,c}$\\
{ $^a$\small Institute of Particle and Nuclear Physics, Henan Normal University, Henan 453007, China}\\
{ $^b$\small School of physics and electronic technology, Liaoning Normal University, Liaoning 116029, China}\\
{ $^c$\small School of physics and electronic engineering, Anyang Normal University, Henan 455000, China}} 
\date{}
\maketitle

\begin{abstract}
Motivated by the rapid development of heavy flavor physics experiment, we study the tree-dominated nonleptonic  $\bar{B}_{u,d,s}^* \to D_{u,d,s}^*V$~($V=D^{*-},D_s^{*-},K^{*-},{\rho}^-$) decays within the factorization approach.   The relevant transition form factors  are calculated by employing the covariant light-front quark model. The helicity amplitudes are calculated and analyzed in detail, and a very clear hierarchy structure $ |H_{-0}| \approx 2|H_{00}| > |H_{0-}|\approx|H_{--}|>|H_{0+}|\approx|H_{++}|$ is presented. The branching fractions are computed and discussed. Numerically, the CKM-favored $\bar{B}^*_q\to D^*_q \rho^{-}$  and $D^*_q D_s^{*-}$ decays have relatively large branching fractions, $\gtrsim {\cal O}(10^{-8})$, and are hopeful to be observed by LHC and Belle-II experiments in the future.
\end{abstract}

\newpage
\section{Introduction}
The $B$ meson weak decay plays an important role in testing the flavor dynamics of the standard model~(SM), searching for the possible hints of new physics, and investigating the approaches of dealing with the hadronic matrix element. Experimentally,   with the successful running of  the $B$ factories, BaBar and Belle,  a lot of $B\bar{B}$ samples have been accumulated, which provides a fertile ground for the b-physics study. Thanks to the ongoing LHCb experiment~\cite{Bediaga:2012py}, many measurements of  $B$ meson decays are refined and some new decay modes are observed. In addition, the running SuperKEKB/Belle-II experiment will also provide us a lot of information about $B$ meson decays~\cite{Abe:2010gxa}. Meanwhile, there are also a plenty of other b-flavored hadron events, such as $\Lambda_b$ and $B^*$ {\it et al.}, will be accumulated in the future, which  may provide much more extensive space for b-physics research.

The $B^*$ meson with quantum number of $n^{2s+1}L_J=1^3S_1$ and $J^P=1^-$ is the vector ground state of $(b\bar{q})$ system~\cite{Tanabashi:2018oca,Isgur:1991wq,Godfrey:1986wj,Eichten:1993ub,Ebert:1997nk}, and can play a similar role as $B$ meson in principle. However, $B^*$ meson decay is dominated by the electromagnetic process $B^*\to B \gamma$, and its weak decay is too rare to be observed in the previous heavy-flavor experiments. Fortunately, due to the rapid development of particle physics experiment in recent years, such situation is  hopeful to be improved by LHC and Belle-II experiments {\it et al.} in the future~\cite{Abe:2010gxa,Bediaga:2012py,Aaij:2014jba}. For instance, the annual integrated luminosity of Belle-II is expected to reach up to $\sim$ 13 $ab^{-1}$ and the $B^*$ weak decays with branching fractions $> {\cal O}(10^{-9})$ are hopeful to be observed~\cite{Abe:2010gxa,Chang:2015jla}. Moreover, the LHC experiment  will also provide a lot of experimental information for $B^*$ weak decays due to the much larger beauty production cross-section of $pp$ collision relative to $e^+e^-$ collision~\cite{Aaij:2010gn}. Therefore, the theoretical studies of $B^*$ weak decays are urgently required for providing some useful suggestions and references to relevant measurements.

Some theoretical studies of $B^*$ weak decays  have been made recently.  In Ref.~\cite{Grinstein:2015aua}, the pure leptonic $B_s^*\to \ell \ell$ and $B_{u,c}^*\to l\nu$ decays are studied, and the detectability of LHC on these decays are analyzed in detail; these decays are revisited in Ref. \cite{Sun:2019xyw} with the decay constant obtained via a relativistic potential model;  in addition, the impact of  $\bar{B}_{s,d}^*\to \mu\mu$ on  $\bar{B}_{s,d}\to \mu\mu$ decays are discussed in Ref.~\cite{Xu:2015eev}, and the authors of Refs.~\cite{Saini:2019tge,Kumbhakar:2018uty,Banerjee:2017upm,Sahoo:2017zke} try to probe the signatures of  physics  beyond the SM through these decays.  Some semileptonic $B^*$ decays are evaluated within the framework of QCD sum rules~\cite{Wang:2012hu,Zeynali:2014wya,Bashiry:2014qia}, Bauer-Stech-Wirbel model~\cite{Chang:2016cdi} and Bethe-Salpeter method~\cite{Wang:2018dtb}; and the effects of new physics are studied in a model independent way in Ref.~\cite{Chang:2018sud}. Some CKM-favored $\bar{B}_q^*\to D_qP$ and $D_qV$ decays are evaluated in the framework of naive factorization~(NF)~\cite{Chang:2015jla,Chang:2015ead}, perturbative QCD~(PQCD)~\cite{Sun:2017xed} and QCD factorization~(QCDF)~\cite{Chang:2016meh,Chang:2016eto}. In this paper, we will pay our attention to the $\bar{B}_q^*\to D_q^* V$~($V=D^{*-}\,,D_s^{*-}\,,K^{*-}\,,{\rho}^-$) decay modes.

Comparing $\bar{B}_q^*\to D_qP$ and $D_qV$ decays with the corresponding $\bar{B}_q\to D_q^*P$ and $D_q^*V$ decays, respectively, one can find a close relationship between them because their main difference is $\bar{B}_q^*\to D_q$ vs. $\bar{B}_q\to D_q^*$~\cite{Chang:2015jla,Chang:2015ead}, and their amplitudes are similar with each other. While, $\bar{B}_q^*\to D_q^* V$ process is  a peculiar decay mode and there is no correspondence from $B$  decays. Comparing with $\bar{B}_q^*\to D_qP$ and $D_qV$ decays, $\bar{B}_q^*\to D_q^* V$ decay is much more complicated because it involves much more allowed helicity states of initial and final mesons contributing to the amplitude, which is worth careful study.  In addition, the form factors of $\bar{B}_q^*\to D_q^*$  and $\bar{B}_q^*\to V$  transitions play an important role in estimating the amplitudes but there is no available results can be used for now; thus, we will calculate these form factors within the framework of a covariant light-front quark model~(CLFQM) in this paper.

Our paper is organized as follows. In section 2, after a brief review of the theoretical framework, the helicity amplitudes of $\bar{B}_q^*\to D_q^* V$ decays are calculated in detail. In section 3, the input parameters used in this work are given, especially,  the relevant form factors are calculated within the CLFQM; after that, the numerical results and discussions for the $\bar{B}_q^*\to D_q^* V$  decays are presented. Finally, we give our summary in section 4.

\section{Theoretical framework}
The effective Hamiltonian responsible for nonleptonic $ \bar{B}^*$ decays can be written as
 \begin{eqnarray}
{\cal H}_{eff} =\frac{G_F}{\sqrt 2}\sum_{p,\,p^\prime=u,\,c}\left[ V_{pb}V^*_{p^\prime q}\sum_{i=1}^2 C_i(\mu)O_i(\mu)+V_{pb}V^*_{p q}\sum_{i=3}^{10} C_i(\mu)O_i(\mu)\right]+h.c.,
\end{eqnarray}
where $G_F$ is the Fermi coupling constant, $ V_{pb}V^*_{p^{(\prime)}q}(q=d,\,s)$ is the product of Cabibbo-Kobayashi-Maskawa~(CKM) matrix elements,  $ C_{i}(\mu)$ is  Wilson coefficient and can be calculated with the perturbation theory~\cite{Buchalla:1995vs,Buras:1998raa}, $O_i$ is local four-quark operator and its explicit form can be found in, for instance, Refs.~\cite{Buchalla:1995vs,Buras:1998raa}.

In order to obtain the decay amplitudes, we have to deal with the hadronic matrix elements of local operators, $\langle V_1V_2|O_i |B^* \rangle$, involved in the amplitude.  A simple way  for this purpose is the naive factorization~(NF) scheme~\cite{Fakirov:1977ta,Cabibbo:1977zv,Bauer:1984zv,Wirbel:1985ji,Bauer:1986bm} based on the color transparency mechanism~\cite{Bjorken:1988kk,Jain:1995dd}. Within the NF  approach, the hadronic matrix element of $B^*\to V_1V_2$ decay can be factorized as
\begin{align}\label{eq:fac}
H_{\lambda_1\lambda_2}^{V_1V_2}\equiv\langle V_1V_2|Q_i|B^*\rangle\simeq\langle V_2|J_2|0\rangle\langle V_1|J_1|B^*\rangle\,,
\end{align}
in which, the recoil vector meson that carries away the spectator quark from $B^*$ meson is called as $V_1$, and the emission one is called as $V_2$; $\lambda_{1(2)}$ is the helicity of $V_{1(2)}$ meson, and the helicity of initial $B^*$ meson satisfies $\lambda_{B^*}=\lambda_{1}-\lambda_{2}$. The two current matrix elements $\langle V_2|J_2|0\rangle$ and $\langle V_1|J_1|B^*\rangle$ in Eq.~\eqref{eq:fac} can be further parameterized  by decay constant and form factors.

In the framework of NF,  the nonfactorizable contributions dominated by the hard gluon exchange are lost. In order to evaluate these QCD corrections to the matrix elements and reduce the scale-dependence, the QCDF approach is explored by BBNS~\cite{Beneke:1999br,Beneke:2000ry}.  In spite of this, the NF approach is employed in this paper due to the following reasons: (i) In the framework of QCDF, the amplitude obtained through NF can be treated as  the leading-order~(LO) contribution of QCDF result. For the $b\to c$ induced tree-dominated nonleptonic $B^{(*)}$ decays,   compared with the LO contribution, the NLO and NNLO nonfactorizable QCD corrections generally give about $4\%$ and $2\%$ contributions~\cite{Beneke:2000ry,Chang:2016eto,Huber:2016xod}, respectively. Therefore,  for the $B^*\to D^* V$ decays studied in this paper, the NF can give relatively reliable predictions, and the small nonfactorizable  QCD correction are numerically trivial before these $B^*$ decay modes are measured precisely.  (ii)  The QCDF approach is not suitable anymore  for the case of heavy emission meson~\cite{Beneke:2000ry}, for instance,  $\bar{B}^*\to D^* \bar{D}^*$ decays.

The decay constant and form factors are essential inputs for evaluating the current matrix elements in Eq.~\eqref{eq:fac}. The former is defined as
\begin{eqnarray}
\langle V_2(\epsilon_2,p_2)|\bar{q}\gamma^{\mu} q|0\rangle =f_{V_2}m_2\epsilon_{2}^{*\mu},
\end{eqnarray}
where $ m_2$ and $ \epsilon_2 $ denote the mass and the polarization vector of $V_2$ meson, respectively. The form factors for $B^*\to V_1$ transition are defined by~\cite{Wang:2007ys}
\begin{align}
{\langle}V_1(\epsilon_1,p_1)|{\bar c}\gamma_{\mu}b|B^*(\epsilon,p){\rangle}
=&\epsilon \cdot {\epsilon_1^*}\left[-P_{\mu}{\tilde V_1(q^2)}+q_{\mu}{\tilde V_2(q^2)}\right]
+\frac{({\epsilon \cdot q})({\epsilon_1^*}\cdot q)}{{m^2_{B^*}}-{m^2_1}}\left[P_{\mu}{\tilde V_3(q^2)}-q_{\mu}{\tilde V_4(q^2)}\right]
 \nonumber \\
&-({\epsilon \cdot q}){\epsilon^*_{1{\mu}}}{\tilde V_5(q^2)}+({\epsilon_1^*}\cdot q)\epsilon_{\mu}{\tilde V_6(q^2)}\,, \label{eq:FFV}\\
{\langle}V_1(\epsilon_1,p_1)|{\bar c}\gamma_{\mu}\gamma_5b|B^*(\epsilon,p){\rangle}
=&-i\varepsilon_{\mu\nu\alpha\beta}\epsilon^{\alpha}{\epsilon_1^{*{\beta}}}\left[P^{\nu}{\tilde A_1(q^2)}-q^{\nu}{\tilde A_2(q^2)}\right] \nonumber \\
&+\frac{2i}{{m^2_{B^*}}-{m^2_1}}\varepsilon_{\mu\nu\alpha\beta}p^{\alpha}{p^{\beta}_1}\left[\epsilon^{\nu}({\epsilon^*_1}\cdot q)\tilde A_3(q^2)-\epsilon_1^{*\nu}(\epsilon \cdot q)\tilde A_4(q^2)\right]\,,\label{eq:FFA}
\end{align}
where $ \epsilon_{0123}=-1 $, $ P=p+p_1$, $ q=p-p_1=p_2$,  and $\epsilon_{(1)}$ is the polarization vector of $ B^*(V_1)$ meson.

Then, after contracting the current matrix elements, we can obtain the $H_{\lambda_1\lambda_2}^{V_1V_2}$ for the $7$ allowed helicity states of final mesons written as
\begin{align}
H_{00}^{V_1V_2}=&f_{V_2}m_{2}   \Big[\frac{p_c(m_{B^*}^2+m_{1}^2-m_{2}^2)}{m_{1}m_{2}}\tilde V_1+\frac{2m_{B^*}^2 p_c^{3} }{(m_{B^*}^2-m_{1}^2)m_{1}m_2}\tilde V_3 -\frac{p_c(m_{B^*}^2-m_{1}^2-m_2^2)}{2 m_{1}m_2}\tilde V_5\nonumber  \\
& +\frac{p_c(m_{B^*}^2-m_{1}^2+m_2^2)}{2 m_{1}m_2}\tilde V_6\Big]\,,\\
H_{++}^{V_1V_2}=&f_{V_2}m_{2}\Big[\frac{3m_{B^*}^2+m_{1}^2-m_2^2}{2m_{B^*}}\tilde A_1 -\frac{m_{B^*}^2-m_{1}^2+m_2^2}{2 m_{B^*}}\tilde A_2 +\frac{2p_c^{2}m_{B^*}}{m_{B^*}^2-m_{1}^2}\tilde A_4-p_c \tilde V_5\Big]\,,  \\
H_{--}^{V_1V_2}=&f_{V_2}m_{2}  \Big[-\frac{3m_{B^*}^2+m_{1}^2-m_2^2}{2m_{B^*}}\tilde A_1 +\frac{m_{B^*}^2-m_{1}^2+m_2^2}{2 m_{B^*}}\tilde A_2 -\frac{2p_c^{2}m_{B^*}}{m_{B^*}^2-m_{1}^2}\tilde A_4-p_c \tilde V_5\Big]\,,   \\
H_{+0}^{V_1V_2}=&f_{V_2}m_{2}  \Big[-\frac{m_{B^*}^2-m_{1}^2}{m_2}\tilde A_1+m_2\tilde A_2 +\frac{2 m_{B^*} p_c }{m_2}\tilde V_1\Big]\,,  \label{eq:Hp0} \\
H_{-0}^{V_1V_2}=&f_{V_2}m_{2} \Big[\frac{m_{B^*}^2-m_{1}^2}{m_2}\tilde A_1-m_2\tilde A_2+\frac{2 m_{B^*} p_c }{m_2}\tilde V_1\Big]\,, \\
H_{0-}^{V_1V_2}=&f_{V_2}m_{2}   \Big[-\frac{m_{B^*}^2+3m_{1}^2-m_2^2 }{2m_{1}}\tilde A_1 +\frac{m_{B^*}^2-m_{1}^2-m_2^2}{2 m_{1}}\tilde A_2-\frac{2p_c^{2} m_{B^*}^2}{(m_{B^*}^2-m_{1}^2) m_{1}}\tilde A_3\nonumber \\
&-\frac{p_c m_{B^*}}{m_{1}} \tilde V_6\Big]\,, \\
H_{0+}^{V_1V_2}=&f_{V_2}m_{2}  \Big[\frac{m_{B^*}^2+3m_{1}^2-m_2^2 }{2m_{1}}\tilde A_1 -\frac{m_{B^*}^2-m_{1}^2-m_2^2}{2 m_{1}}\tilde A_2+\frac{2p_c^{2} m_{B^*}^2}{(m_{B^*}^2-m_{1}^2) m_{1}}\tilde A_3\nonumber \\
&-\frac{p_c m_{B^*}}{m_{1}} \tilde V_6\Big]\,,
\end{align}
where, $p_c=\frac{\sqrt{[m_{B^*}^2-(m_{1}+m_{2})^2][m_{B^*}^2-(m_{1}-m_{2})^2]}}{2m_{B^*}}$.

Using the formulas given above, we can finally obtain the helicity amplitudes of tree-dominated ${\bar{B}_{u,d,s}^*} \to D_{u,d,s}^{*}V$ $(V=D^{*-},{D}_s^{*-},{K}^{*-}, \rho^-) $ decays, which can be written as
\begin{align}
{\cal A}({B^{*-}\to D^{*0}K^{*-}})=&\frac{G_F}{\sqrt 2}[H_{\lambda_{D^*}\lambda_{K^{*}}}^{D^*K^{*}}V_{cb}V_{us}^* \alpha_1+ H_{\lambda_{K^{*}}\lambda_{D^*}}^{K^{*}D^*}V_{cb}V_{us}^*\alpha_2 ]\,,\\
{\cal A}({B^{*-}\to D^{*0}{\rho}^-})=&\frac{G_F}{\sqrt 2}[H_{\lambda_{D^*}\lambda_{\rho}}^{D^*\rho} V_{cb}V_{ud}^* \alpha_1+ H_{\lambda_{\rho}\lambda_{D^*}}^{\rho D^*} V_{cb}V_{ud}^*\alpha_2 ]\,, \\
{\cal A}({B^{*-}\to D^{*0}D^{*-}})=&\frac{G_F}{\sqrt 2}H_{\lambda_{D^{*0}}\lambda_{D^{*-}}}^{D^{*0}D^{*-}}[V_{cb}V_{cd}^* (\alpha_1+\alpha_4+\alpha_{4,EW} )+V_{ub}V_{ud}^*(\alpha_4+\alpha_{4,EW})]\,,\\
{\cal A}({B^{*-}\to D^{*0}D_s^{*-}})=&\frac{G_F}{\sqrt 2}H_{\lambda_{D^{*}}\lambda_{D^{*}_s}}^{D^{*}D^{*}_s}[V_{cb}V_{cs}^* (\alpha_1+\alpha_4+\alpha_{4,EW} )+V_{ub}V_{us}^*(\alpha_4+\alpha_{4,EW})]\,,\\
{\cal A}({\bar B^{*0}\to D^{*+}K^{*-}})=&\frac{G_F}{\sqrt 2}H_{\lambda_{D^*}\lambda_{K^*}}^{D^*K^*}V_{cb}V_{us}^* \alpha_1\,,\\
{\cal A}({\bar B^{*0}\to D^{*+}\rho^-})=&\frac{G_F}{\sqrt 2}H_{\lambda_{D^*}\lambda_{\rho}}^{D^*\rho}V_{cb}V_{ud}^* \alpha_1\,, \\
{\cal A}({\bar B^{*0}\to D^{*+}D^{*-}})=&\frac{G_F}{\sqrt 2}H_{\lambda_{D^{*+}}\lambda_{D^{*-}}}^{D^{*+}D^{*-}}[V_{cb}V_{cd}^* (\alpha_1+\alpha_4+\alpha_{4,EW} )+V_{ub}V_{ud}^*(\alpha_4+\alpha_{4,EW})]\,,\\
{\cal A}({\bar B^{*0}\to D^{*+}D_s^{*-}})=&\frac{G_F}{\sqrt 2}H_{\lambda_{D^{*}}\lambda_{D^{*}_s}}^{D^{*}D^{*}_s}[V_{cb}V_{cs}^* (\alpha_1+\alpha_4+\alpha_{4,EW} )+V_{ub}V_{us}^*(\alpha_4+\alpha_{4,EW})]\,,\\
{\cal A}({\bar B^{*0}_s \to D_s^{*+}K^{*-}})=&\frac{G_F}{\sqrt 2}H_{\lambda_{D_s^{*}K^{*}}}^{D_s^{*}K^{*}}V_{cb}V_{us}^* \alpha_1\,, \\
{\cal A}({\bar B^{*0}_s\to D^{*+}_s {\rho}^-})=&\frac{G_F}{\sqrt 2}H_{\lambda_{D^*_s}\lambda_{\rho}}^{D^*_s\rho}V_{cb}V_{ud}^* \alpha_1\,,\\
{\cal A}({\bar B_s^{*0}\to D_s^{*+}D^{*-}})=&\frac{G_F}{\sqrt 2}H_{\lambda_{D^{*}_s}\lambda_{D^{*}}}^{D^{*}_sD^{*}}[V_{cb}V_{cd}^* (\alpha_1+\alpha_4+\alpha_{4,EW} )+V_{ub}V_{ud}^*(\alpha_4+\alpha_{4,EW})]\,, \\
{\cal A}({\bar B_s^{*0}\to D_s^{*+}D_s^{*-}})=&\frac{G_F}{\sqrt 2}H_{\lambda_{D_s^{*+}}\lambda_{D_s^{*-}}}^{D_s^{*+}D^{*-}_s}[V_{cb}V_{cs}^* (\alpha_1+\alpha_4+\alpha_{4,EW} )+V_{ub}V_{us}^*(\alpha_4+\alpha_{4,EW})]\,,
\end{align}
where, $\alpha_1=C_1+\frac{C_2}{N_c}$, $\alpha_2=C_2+\frac{C_1}{N_c}$, $\alpha_4=C_4+\frac{C_3}{N_c}$ and $\alpha_{4,EW}=C_{10}+\frac{C_9}{N_c}$ are effective coefficients, and $N_c=3$ denotes the number of colors. 

 Using the helicity amplitudes given above, one can further obtain the branching fraction of $ B^*\to D^* V $ decay defined as
\begin{eqnarray}
{\cal B}(B^* \to D^* V)=\frac{1}{3} \frac{1}{8\pi} \frac{p_c}{m_{B^*}^2 {\Gamma}_{\rm tot}(B^*)} \sum_{\lambda_{B^*}\lambda_{D^*}\lambda_{V}}|{\cal A}(B^* \to D^* V)|^2,
\end{eqnarray}
where $ {\Gamma}_{\rm tot}(B^*) $ is the total decay width of $ B^* $ meson,  and the factor $1/3 $ is caused by averaging over the spins of initial state.

\section{Numerical Results and Discussions}
Using the theoretical formulas given in the last section, we then present our numerical evaluation and discussions. Firstly, we would like to clarify the values of inputs used in our numerical calculation. For the well-known Fermi coupling constant $G_F$ and the masses of mesons, we take their central values given by PDG~\cite{Tanabashi:2018oca}. For the CKM matrix elements, we adopt the Wolfenstein parameterization, and  the four parameters $A$, $\lambda$, $\rho$ and $\eta$ are~\cite{Tanabashi:2018oca}
\begin{equation}
A={0.836}^{+0.015}_{-0.015}\,, \quad
\lambda={0.22453}^{+0.00044}_{-0.00044}\,, \quad
\overline{\rho}={0.122}^{+0.018}_{-0.017}\,, \quad
\overline{\eta}={0.355}^{+0.012}_{-0.011}\,.
\end{equation}
Using these inputs, we can easily obtain the values of CKM elements relevant to this work that $V_{ud}=0.97448^{+0.00010}_{-0.00010}$, $V_{us}={0.22453}^{+0.00044}_{-0.00044}$, $V_{ub}=0.00122^{+0.00018}_{-0.00017}-i\, 0.00354^{+0.00014}_{-0.00013}$, $V_{cd}=-0.22438^{+0.00044}_{-0.00044}-i\, 0.00015^{+0.00001}_{-0.00001}$, $V_{cs}=0.97359^{+0.00010}_{-0.00010}$ and $V_{cb}=0.04215^{+0.00077}_{-0.00077} $  at the level of ${\cal O}(\lambda^5)$.
For the decay constants of emission mesons, we take their values extracted from experiment data and predicted by Lattice QCD
\begin{eqnarray}
&&f_{D^{*}}=223.5\pm8.4\,{\rm MeV}~\text{\cite{Lubicz:2016bbi}}\,, \quad  f_{D_s^*}=268.8\pm6.6\, {\rm MeV}~\text{\cite{Lubicz:2016bbi}}\,,\nonumber\\
  &&f_{K^*}=204\pm 7 \,{\rm MeV}~\text{\cite{Straub:2015ica}}\,, \quad f_{\rho}=210\pm4 \,{\rm MeV}~\text{\cite{Braun:2016wnx}}\,.
 \end{eqnarray}
 The total decay width of $ B^* $ meson is the essential input for evaluating the branching fraction, but there is no available experimental result for now. Based on the fact that the radiative  process $B^*\to B\gamma$ dominates the decays of $B^*$ meson~\cite{Tanabashi:2018oca}, we can take the approximation that $\Gamma_{\rm tot}(B^*)\simeq \Gamma(B^*\to B\gamma)$. The predictions for $\Gamma(B^*\to B\gamma)$ have been obtained in various theoretical models~\cite{Goity:2000dk,Ebert:2002xz,Zhu:1996qy,Aliev:1995zlh,Colangelo:1993zq,Choi:2007se,Cheung:2014cka}.  In this paper, the light-front quark model~(LFQM) is employed to evaluate $\Gamma(B^*\to B\gamma)$. The relevant theoretical formulas have been obtained in Ref.~\cite{Choi:2007se}. Using the values of Gaussian parameter $\beta$ given in Refs.~\cite{Chang:2018zjq,Chang:2019mmh}, we can obtain the updated LFQM predictions for  $\Gamma(B^*\to B\gamma)$ as follows
\begin{align}
\label{eq:GtotBu}
\Gamma_{\rm{tot}}(B^{*-})&\simeq \Gamma(B^{*-}\to B^- \gamma)=(349\pm{18})\,{\rm eV},\\
 \label{eq:GtotBd}
\Gamma_{\rm{tot}}(\bar{B}^{*0})&\simeq \Gamma(\bar{B}^{*0}\to \bar{B}^0 \gamma)=(116\pm6)\,{\rm eV},\\
 \label{eq:GtotBs}
\Gamma_{\rm{tot}}(\bar{B}^{*}_s)&\simeq \Gamma(\bar{B}^{*}_s\to \bar{B}_s \gamma)=(84^{+11}_{-9})\,{\rm eV},
\end{align}
which agree with the ones obtained in the previous works~\cite{Goity:2000dk,Ebert:2002xz,Zhu:1996qy,Aliev:1995zlh,Colangelo:1993zq,Choi:2007se,Cheung:2014cka}.

\begin{table}[t]
\caption{The form factors of ${B}^{*}\to (D^{*},\,K^*,\,\rho) $ and  ${B}^{*}_{s}\to D^{*}_s$ transitions in the CLFQM.}
\vspace{-0.1cm}\footnotesize
\begin{center}\setlength{\tabcolsep}{5pt}
\begin{tabular}{l|ccc|ccc|ccc|cccc}
\hline\hline
  &$F^{B^* \to D^*}(0)$&a&b     &$F^{B^* \to K^*}(0)$&a&b      &$F^{B^* \to \rho}(0)$&a&b  &$F^{B_s^* \to D_s^*}(0)$&a&b\\
\hline
$\tilde A_1$&$0.66$& $1.31$& $0.42$ &$0.33$&$1.75$&$0.89$  &$0.27$&$1.79$&$0.97$ &$0.65$&$1.42$&$0.64$\\
$\tilde A_2$&$0.35$& $1.32$& $0.42$ &$0.27$&$1.75$&$0.88$  &$0.25$&$1.80$&$0.97$ &$0.38$&$1.47$&$0.67$\\
$\tilde A_3$&$0.07$& $1.79$& $1.10$ &$0.07$&$2.28$&$2.20$  &$0.07$&$2.39$&$2.37$ &$0.10$&$1.89$&$1.33$\\
$\tilde A_4$&$0.08$& $1.81$& $1.15$ &$0.07$&$2.29$&$2.33$  &$0.06$&$2.35$&$2.46$ &$0.09$&$1.88$&$1.36$\\
$\tilde V_1$&$0.67$& $1.31$& $0.43$ &$0.33$&$1.74$&$0.96$  &$0.28$&$1.79$&$0.01$ &$0.66$&$1.43$&$0.64$\\
$\tilde V_2$&$0.36$& $1.32$& $0.42$ &$0.27$&$1.74$&$0.95$  &$0.25$&$1.80$&$1.02$ &$0.38$&$1.48$&$0.67$\\
$\tilde V_3$&$0.13$& $1.72$& $1.01$ &$0.11$&$2.16$&$2.04$  &$0.11$&$2.23$&$2.16$ &$0.15$&$1.79$&$1.20$\\
$\tilde V_4$&$0.00$&$-0.08$&$1.24$  &$-0.01$&$2.91$&$4.24$ &$-0.03$&$2.77$&$3.74$ &$-0.02$&$2.22$&$1.92$\\
$\tilde V_5$&$1.17$& $1.30$& $0.40$ &$0.68$&$1.71$&$0.90$  &$0.60$&$1.76$&$0.95$ &$1.19$&$1.41$&$0.61$\\
$\tilde V_6$&$0.48$& $1.29$& $0.40$ &$0.16$&$1.67$&$0.81$  &$0.14$&$1.70$&$0.82$ &$0.53$&$1.35$&$0.56$\\
\hline\hline
\end{tabular}
\end{center}
\label{tab:BstarVPVV}
\end{table}
Besides the input parameters given above, the transition form factors $ \tilde V_{1-6}^{B^*\to V_1}(m_2^2)$, $\tilde A_{1- 4}^{B^*\to V_1}(m_2^2)$ are  also essential ingredients for the estimation of a certain nonleptonic $B^*$ decay. However, there is no available result until now. In this work, we adopt the CLFQM~\cite{Jaus:1999zv,Jaus:2002sv,Cheng:2003sm}  to evaluate their values.   Our theoretical results for $\tilde V_{1-6}(q^2)$ and $\tilde A_{1-4}(q^2)$ defined by Eqs.~\eqref{eq:FFV} and \eqref{eq:FFA} are given explicitly in the appendix.  It should be noted that the convenient Drell-Tan-West frame, $q^{+}=0$, is used in the CLFQM~\cite{Jaus:1999zv,Jaus:2002sv,Cheng:2003sm}. It implies that the form factors are known only for space-like momentum transfer, since $q^2=-{q}_\bot^2\leqslant 0$, by using the formulas given  in the appendix, and the ones in the time-like region need  an additional $q^2$ extrapolation. The momentum dependences of form factors in the space-like region can be well parameterized and reproduced via the three parameter form~(dipole approximation),
\begin{eqnarray}\label{eq:dipo}
F(q^2)=\frac{F(0)}{1-a\,(q^2/m^2_{B^*})+b\,(q^2/m^2_{B^*})^2}\,,
\end{eqnarray}
 where, $F= \tilde V_{1-6}$ and $\tilde A_{1- 4}$. The parameters $a$, $b$ and $F(0)$ can be firstly determined  in the space-like region; and then, we employ these results to evaluate $F(q^2)$ at $q^2\geqslant 0$ via Eq.~\eqref{eq:dipo}. Using the best-fit values of constituent quark masses and Gaussian parameter  obtained in Refs.~\cite{Chang:2018zjq,Chang:2019mmh}, we give our numerical results for  the form factors of ${B}^{*}\to (D^{*},\,K^*,\,\rho) $ and  ${B}^{*}_{s}\to D^{*}_s$ transitions in Table~\ref{tab:BstarVPVV}. In the following numerical calculation, these values and $10\%$ of them are treated as default inputs and uncertainties, respectively.

\begin{table}[t]
\caption{The branching fractions and helicity fractions ($\%$) of $\bar{B}^{*}_q\to D^{*}_q V $ decays.}
\vspace{-0.0cm}
\begin{center}\setlength{\tabcolsep}{8pt}
\begin{tabular}{l|c|cccccccc}
\hline\hline
Decay mode        &${\cal B}\,$  &$f_{00} $&$f_{--}$&$f_{++}$&$f_{-0}$&$f_{+0}$&$f_{0-}$&$f_{0+}$\\\hline
${B}^{*-} \to D^{*0} K^{*-}$         &  $1.10^{+0.01+0.19}_{-0.01-0.17}\times10^{-9}$
& 24.4  &4.5&0.3    &69.2&0.0    &1.4  &0.2  \\
${B}^{*-} \to D^{*0} \rho^{-}$       & $2.23^{+0.04+0.39}_{-0.04-0.35}\times10^{-8}$
&24.1  &3.3&0.2    &71.0&0.0    &1.2  &0.2  \\
${B}^{*-} \to D^{*0} D^{*-}$       & $1.44^{+0.11+0.24}_{-0.11-0.22}\times10^{-9}$
&12.9   &13.1&1.8   &56.6&0.4    &13.8  &1.5 \\
${B}^{*-} \to D^{*0} D_s^{*-}$    & $3.71^{+0.18+0.64}_{-0.18-0.57}\times10^{-8}$
&12.1  &14.1&2.0    &54.8 &0.5   &14.7  &1.8 \\
$\bar{B}^{*0} \to D^{*+} K^{*-}$  & $3.40^{+0.24+0.58}_{-0.23-0.52}\times10^{-9}$
&19.7  &3.2 &0.3   &73.2&0.0   &3.4  &0.2 \\
$\bar{B}^{*0} \to D^{*+} \rho^-$  & $6.85^{+0.26+1.17}_{-0.26-1.05}\times10^{-8}$
&20.1  &2.5  &0.2  &74.4 &0.0   &2.6  &0.2 \\
$\bar{B}^{*0} \to D^{*+} D^{*-}$   & $4.33^{+0.33+0.74}_{-0.32-0.66}\times10^{-9}$
&12.9   &13.1  &1.8 &56.6 &0.4   &13.8 &1.5\\
$\bar{B}^{*0} \to D^{*+} D_s^{*-}$ & $1.11^{+0.06+0.19}_{-0.05-0.17}\times10^{-7}$
&12.1  &14.1 &2.0   &54.8 &0.5   &14.7  &1.8  \\
$\bar{B}_s^{*0} \to D_s^{*+} K^{*-}$ & $4.80^{+0.34+0.83}_{-0.32-0.74}\times10^{-9}$
&20.2  &3.2  &0.3  &72.7  &0.0  &3.5  &0.2 \\
$\bar{B}_s^{*0} \to D_s^{*+} \rho^-$ & $9.39^{+0.36+1.63}_{-0.35-1.46}\times10^{-8}$
&20.4  &2.5 &0.2   &74.1  &0.0  &2.7  &0.2   \\
$\bar{B}_s^{*0} \to D_s^{*+} D^{*-}$& $6.10^{+0.47+1.03}_{-0.45-0.92}\times10^{-9}$
&13.3  &13.0 &1.7   &56.2  &0.3  &14.1  &1.4  \\
$\bar{B}_s^{*0} \to D_s^{*+} D_s^{*-}$ & $1.54^{+0.08+0.26}_{-0.07-0.24}\times10^{-7}$
&12.9  &13.9  &1.9  &54.4 &0.4   &15.1  &1.5 \\
\hline\hline
\end{tabular}
\end{center}
\label{tab:BstarDstarV}
\end{table}

\begin{table}[t]
\caption{Helicity diagrams for the helicity states of $\bar{B}^*\to V_1 V_2$ decay, $(\lambda_{1},\lambda_{2})$. Initial $B^*$ meson is at rest and appears at top left of diagrams.   S~(F) denotes the corresponding contribution of helicity state is  suppressed~(favored) by $(V-A)$ interaction and/or spin flip. See text for further explanation. }
\vspace{0cm}
\begin{center}\setlength{\tabcolsep}{15pt}
\begin{tabular}{l|ccccccccc}
\hline\hline
Helicity state   &$(0,0)_1$  &$(0,0)_2$&$(-,-)$&$(+,+)$
\\\hline
Helicity diagram &\makecell[c]{ \includegraphics[width=1.5cm]{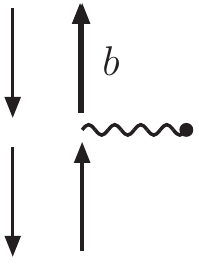}}&\makecell[c]{ \includegraphics[width=1.5cm]{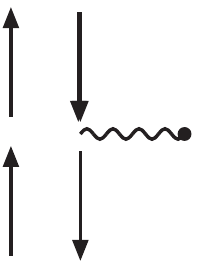}} &\makecell[c]{ \includegraphics[width=1.5cm]{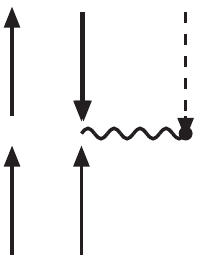}} &\makecell[c]{ \includegraphics[width=1.5cm]{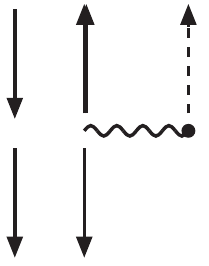}}
\\\hline
$(V-A)$/spin flip &  F/F  &S/F&F/S&S/S
\\\hline\hline
Helicity state  &$(-,0)$  &$(+,0)$&$(0,-)$&$(0,+)$
\\\hline
Helicity diagram &\makecell[c]{ \includegraphics[width=1.5cm]{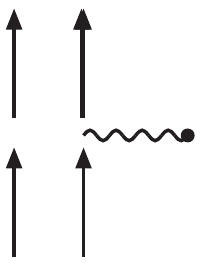}}&\makecell[c]{ \includegraphics[width=1.5cm]{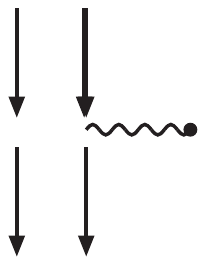}} &\makecell[c]{ \includegraphics[width=1.5cm]{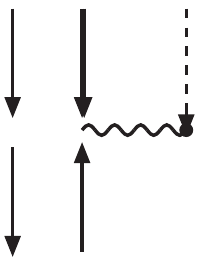}} &\makecell[c]{ \includegraphics[width=1.5cm]{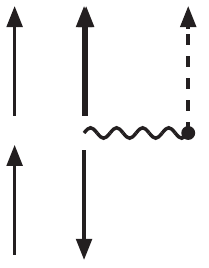}}
\\\hline
$(V-A)$/spin flip &  F/F  &S/F&F/S&S/S
\\\hline\hline
\end{tabular}
\end{center}
\label{tab:ha}
\end{table}

Using the theoretical formulas given in the last section and inputs given above, we then present our predictions for the branching fractions of $\bar{B}^{*}_q\to D^{*}_q V $ decays in Table~{\ref{tab:BstarDstarV}, where the first theoretical error is caused by the uncertainties of  CKM parameters, decay constants and total decay width, and the second one is caused by the form factors. Besides, in order to clearly show the relative strength of each helicity amplitude, we also list the numerical results for the helicity fraction defined as 
\begin{equation}
f_{\lambda_1\lambda_2}(\bar{B}^{*}_q\to D^{*}_q V )=\frac{\left|{\cal A}_{\lambda_1\lambda_2}(\bar{B}^{*}_q\to D^{*}_q V)\right|^2}{\sum_{\lambda_1,\lambda_2}\left|{\cal A}_{\lambda_1\lambda_2}(\bar{B}^{*}_q\to D^{*}_q V)\right|^2}
\end{equation} 
in Table~\ref{tab:BstarDstarV}. The following are some discussions.
\begin{itemize}
\item From Table~\ref{tab:BstarDstarV}, one can find a very clear hierarchy of the branching fractions that ${\cal B}(\bar{B}^*_q\to D^*_q\rho^-)>{\cal B}(\bar{B}^*_q\to D^*_qK^{*-})$ and ${\cal B}(\bar{B}^*_q\to D^*_qD_s^{*-})>{\cal B}(\bar{B}^*_q\to D^*_qD^{*-})$, which is mainly  caused by the  CKM factors that $V_{cb}V_{ud}:V_{cb}V_{us}\approx V_{cb}V_{cs}:V_{cb}V_{cd} \approx 1/\lambda$; meanwhile, ${\cal B}(\bar{B}^*_q\to D^*_qD_s^{*-})>{\cal B}(\bar{B}^*_q\to D^*_q\rho^-)$ and ${\cal B}(\bar{B}^*_q\to D^*_qD^{*-})>{\cal B}(\bar{B}^*_q\to D^*_qK^{*-})$ because $f_{D_s^*}>f_\rho$ and $f_{D^*}>f_{K^*}$, respectively. The CKM favored $\bar{B}^*_q\to D^*_q \rho^{-}$  and $D^*_q \bar{D}_s^*$ decays have relatively large branching fractions, $\gtrsim {\cal O}(10^{-8})$, and therefore are hopeful to be observed by LHC and Belle-II experiments. 

In addition, the $\bar{B}^*\to V_LV_L $~($V_L$ denotes light vector meson) decay modes should have much smaller branching fraction, $< {\cal O}(10^{-9})$, since they are suppressed  at least by the CKM factor and relatively small form factors. They are generally out of the scope of  LHC and Belle-II experiments, and thus are not considered in this paper.

\item The SU(3) flavor symmetry acting on the spectator quark requires that
\begin{align}
{\cal A}({B}^{*-}\to D^{*0}V)\approx{\cal A}(\bar{B}^{*0}_d\to D^{*+}V)\approx{\cal A}(\bar{B}^{*0}_s\to D^{*+}_sV)\,,
\end{align}
which implies the relation that
\begin{align}
{\cal B}({B}^{*-}\to D^{*0}V):{\cal B}(\bar{B}^{*0}_d\to D^{*+}V):{\cal B}(\bar{B}^{*0}_s\to D^{*+}_sV)\approx \frac{1}{\Gamma_{\rm{tot}}(B^{*-})}: \frac{1}{\Gamma_{\rm{tot}}(\bar{B}^{*0})}:\frac{1}{\Gamma_{\rm{tot}}(\bar{B}^{*}_s)}\,.
\end{align}
From Eqs.~(\ref{eq:GtotBu}-\ref{eq:GtotBs}) and Table~\ref{tab:BstarDstarV}, it can be easily found that our  numerical results  agree well with such relation required by the SU(3) flavor symmetry.
\item There is also a clear hierarchy of helicity amplitudes for a given $\bar{B}^{*}_q\to D^{*}_qV$ decay. The helicity picture for the case of $\lambda_{B^{*}}=0$ is similar to the case of  $\bar{B}_q\to D^{*}_qV$ decay~\cite{Ali:1978kn,Korner:1979ci,Kagan:2004uw}, the only difference is the helicity of spectator quark. As shown in Table~\ref{tab:ha}, relative to the $(\lambda_{D^{*}},\lambda_{V})=(0,0)$ helicity state, the contribution of $(-,-)$ state, $H_{--}$, is suppressed since the $b$ quark has to flip its spin in the interaction; for the  contribution of  $(+,+)$ state,  besides of spin flip, it is also suppressed by the $(V-A)$ interaction since the final quark in the  $(V-A)$ interaction appears in the ``wrong'' helicity.  Therefore,  the helicity amplitudes, $H_{00}$, $H_{--}$ and $H_{++}$, should satisfy the relation
\begin{align}
|H_{00}|>|H_{--}|>|H_{++}|\,.
\end{align}

More explicitly, for the case of light $V$ meson, the relation $|H_{00}|:|H_{--}|:|H_{++}|\approx 1:2m_V/m_{B^*}:2m_Vm_{D^*}/m_{B^*}^2$ expected in the $\bar{B}_q\to D^{*}_qV_L$ decay~\cite{Ali:1978kn,Korner:1979ci,Kagan:2004uw} is also satisfied by the $\bar{B}^{*}_q\to D^{*}_qV_L$ decay. For the case of the heavy $V$ meson, the suppression caused by the spin flip is not as strong as the case of light $V$ meson, therefore the $f_{00}$ is relatively small. Our numerical results in Table~\ref{tab:BstarDstarV} are consistent with the analyses mentioned above.

The similar analyses can be further applied to the cases of  $\lambda_{B^{*}}=-$ and $+$. As a result, it is expected that $|H_{-0}|>|H_{0+}|$ and  $|H_{+0}|\gtrsim|H_{0-}|$. However, the later is not satisfied numerically even though they  follow $|H_{+0}|:|H_{0-}|\approx m_{D^{*}}/m_{V}$ in  form. It is caused by the fact that the main contributions related to $\tilde V_1$ and $\tilde A_1$ in $H_{+0}$, Eq.~\eqref{eq:Hp0}, almost completely cancel each other out since $(\tilde V_1-\tilde A_1)\lesssim  {\cal O}(10^{-2})$  predicted by CLFQM. 

\item After making some comparisons on the helicity states: $(0,0)$~vs.~$(-,0)$, $(-,-)$~vs.~$(0,-)$ and $(+,+)$~vs.~$(0,+)$  in Table~\ref{tab:ha}, we find that their helicity diagrams are the same except for the helicity of spectator quark which is trivial for analyzing the suppressions induced by $(V-A)$ interaction and spin flip. Therefore, it is expected that
 \begin{align} \label{eq:rela2}
|H_{-0}|\approx2|H_{00}|\,,\quad |H_{0-}|\approx|H_{--}|\,,\quad |H_{0+}|\approx|H_{++}|\,.
\end{align}
The factor $2$ in the first relation is caused by the following reason:  the vector state $|J,J_z\rangle=|1,0\rangle$ can be expanded in terms of its constituent (anti-)quark's spin states as $|1,0\rangle=\frac{1}{\sqrt{2}}(|\frac{1}{2},-\frac{1}{2}\rangle|\frac{1}{2},\frac{1}{2}\rangle+|\frac{1}{2},\frac{1}{2}\rangle|\frac{1}{2},-\frac{1}{2}\rangle)$, in which the first and the second term correspond to the $B^*$ meson, as well as the recoil vector meson, in $(0,0)_1$ and $(0,0)_2$ states (see  Table~\ref{tab:ha}), respectively; while, for the $B^*$ and recoil vector mesons in $(-,0)$ helicity state, we have $|1,-1\rangle=|\frac{1}{2},-\frac{1}{2}\rangle|\frac{1}{2},-\frac{1}{2}\rangle$. Therefore, the contribution of $(0,0)\approx(0,0)_1$ helicity state receives an additional factor $1/2$ relative to the  contribution of  $(-,0)$ state. The effect of such normalization factor results in a significant difference between $B^{*} \to VV$ and $B \to VV$ decay modes that the former is dominated by $(-,0)$ state but the later is dominated by $(0,0)$ state.

The findings given by Eq.~\eqref{eq:rela2} can be easily confirmed by our numerical  results  listed in Table \ref{tab:BstarDstarV}. Taking $\bar{B}^{*0} \to D^{*+} K^{*-}$ decay as an example, we obtain
 \begin{align}
|H_{-0}|:|H_{00}|=1.93\;{\rm vs.}\;2\,,\quad |H_{0-}|:|H_{--}|=1.03\;{\rm vs.}\;1\,,\quad |H_{0+}|:|H_{++}|=0.89\;{\rm vs.}\;1\,,
\end{align}
where, for the two values in each relation, the former is our numerical result and the later is the expectation of Eq.~\eqref{eq:rela2}.

Combining the findings given above, we can finally conclude the hierarchy of contributions of helicity states  as
 \begin{align} \label{eq:rela3}
|H_{-0}| \approx 2|H_{00}| > |H_{0-}|\approx|H_{--}|>|H_{0+}|\approx|H_{++}|\,.
\end{align}
\end{itemize}

\section{Summary}
In this paper, motivated by the experiments of heavy flavor physics  at running LHC and  SuperKEKB/Belle-II with high-luminosity, the tree-dominated nonleptonic  $\bar{B}_q^* \to D_{q}^*V$~( $q=u\,,d\,,s$ and $V=D^{*-},D_s^{*-},K^{*-},{\rho}^-$) decays are studied first within the framework of  factorization approach, in which the transition form factors of  $\bar{B}_q^* \to D_{q}^*$ and $\bar{B}^* \to K^*\,,\rho$ transitions are calculated within the covariant light-front quark model. The helicity amplitudes are calculated and analyzed in detail. It is found that these decays are dominated by the $(\lambda_{ D_{q}^*},\lambda_{V})=(-,0)$ helicity state, and the contribution of $(0,0)$ state is about half of  the one of $(-,0)$ state in amplitude. It is obviously different from the $B$ meson decay, which is dominated by the $(0,0)$ state.  Moreover, the helicity amplitudes of $\bar{B}_q^* \to D_{q}^*V$ decays follow a very clear hierarchy structure given by Eq.~\eqref{eq:rela3}. The branching fractions are computed, and the effects of CKM factor, SU(3) flavor symmetry and total decay width are discussed in detail. Numerically, the CKM-favored $\bar{B}^*_q\to D^*_q \rho^{-}$  and $D^*_q D_s^{*-}$ decays have relatively large branching fractions, $\gtrsim {\cal O}(10^{-8})$, and are hopeful to be observed by LHC and Belle-II experiments in the future.

\begin{appendix}
\section*{Appendix: the form factors of $V'\to V''$ transition in the CLFQM}
The light-front quark model provides a conceptually simple but practically feasible framework for calculating the non-perturbative quantities, and  has been used extensively to study the weak decays of hadrons (for instance, see Refs.~\cite{Jaus:1996np,Barik:1997qq,Chang:2017sdl,Wang:2017mqp,Choi:2009ai,Choi:2009ym,Choi:2010ha,Choi:2010zb,Hwang:2001wu,Wang:2008ci,Wang:2008xt,Shen:2008zzb,Wang:2009mi,Cheng:2017pcq,Kang:2018jzg}). Using the theoretical formalism of the CLFQM detailed in Refs.~\cite{Jaus:1999zv,Jaus:2002sv,Cheng:2003sm}, we  obtain the form factors of $V'\to V''$ transition written as
 \begin{align}
\tilde A_1(q^2)=&\frac{N_c}{16\pi^3}\int dxd^2k_\perp\frac{h'h''}
{\bar{x}\hat{N}_1^\prime\hat{N}_1^{\prime\prime}}(-4)\Big[-2A_1^{(2)}
+\frac{1}{4}\Big({m'_1}^2+{m''_1}^2-q^2+\hat{N}_1^\prime+\hat{N}_1^{\prime\prime}\Big)
-\frac{1}{2}{m'_1}{m''_1}\nonumber\\
+&A_1^{(1)}\Big(m^2_2-\frac{{M'}^2+{M''}^2}{2}+\frac{1}{2}q^2+{m'_1}{m''_1}-{m'_1}m_2-{m''_1}{m_2}\Big)
\nonumber\\
+&\Big(\frac{1}{D_{V'}}+\frac{1}{D_{V''}}\Big)({m'_1}+{m''_1})A_1^{(2)}\Big]\,,\\
\tilde  A_2(q^2)=&\frac{N_c}{16\pi^3}\int dxd^2k_\perp\frac{h'h''}
{\bar{x}\hat{N}_1^\prime\hat{N}_1^{\prime\prime}}4\Big[{m'_1}{m_2}-\frac{1}{2}{m'_1}{m''_1}-\frac{1}{4}(
{m'_1}^2+{m''_1}^2-q^2+\hat{N}_1^\prime+\hat{N}_1^{\prime\prime}
)\nonumber\\
+&\frac{x}{2}({M'}^2+{M''}^2-q^2)-\frac{k_\bot\cdot q_\bot}{2q^2}({M'}^2-{M''}^2-q^2)+A_1^{(2)}\frac{{M'}^2-{M''}^2}{q^2}
\nonumber\\
+&A_2^{(1)}\Big(
m^2_2-\frac{{M'}^2+{M''}^2}{2}+Z_2
+\frac{1}{2}q^2+{m'_1}{m''_1}-{m'_1}{m_2}-{m''_1}{m_2}\Big)
\nonumber\\
+&\Big(\frac{-{m'_1}+{m''_1}-2{m_2}}{D_{V'}}+\frac{-{m'_1}+{m''_1}+2{m_2}}{D_{V''}}\Big)A_1^{(2)}\Big]\,,\\
\tilde A_3(q^2)=&\frac{N_c}{16\pi^3}\int dxd^2k_\perp\frac{h'h''}
{\bar{x}\hat{N}_1^\prime\hat{N}_1^{\prime\prime}}(-4)({M'}^2-{M''}^2)\Big[
\big(A_4^{(2)}+A_1^{(1)}-A_2^{(2)}-A_2^{(1)}\big)\nonumber\\
+&\frac{{m'_1}}{D_{V''}}\big(-A_4^{(2)}-2A_3^{(2)}-A_2^{(2)}+2A_2^{(1)}+2A_1^{(1)}-1\big)
+\frac{{m''_1}}{D_{V''}}\big(-A_4^{(2)}+A_2^{(2)}+A_2^{(1)}-A_1^{(1)}\big)\nonumber\\
+&\frac{2{m_2}}{D_{V''}}\big(A_3^{(2)}+A_2^{(2)}-A_1^{(1)}\big)
+\frac{2}{D_{V'}D_{V''}}\big(-A_2^{(3)}-A_1^{(3)}+A_1^{(2)}\big)\Big]\,,
\\
\tilde A_4(q^2)=&\frac{N_c}{16\pi^3}\int dxd^2k_\perp\frac{h'h''}
{\bar{x}\hat{N}_1^\prime\hat{N}_1^{\prime\prime}}4({M'}^2-{M''}^2)\Big[
\big(-A_4^{(2)}-A_1^{(1)}+A_2^{(2)}+A_2^{(1)}\big)\nonumber\\
+&\frac{{m'_1}}{D_{V'}}\big(A_4^{(2)}-A_2^{(2)}-A_2^{(1)}+A_1^{(1)}\big)
+\frac{{m''_1}}{D_{V'}}\big(A_4^{(2)}-2A_3^{(2)}+A_2^{(2)}\big)
+\frac{2{m_2}}{D_{V'}}\big(A_3^{(2)}-A_2^{(2)}\big)\nonumber\\
+&\frac{2}{D_{V'}D_{V''}}\big(A_1^{(3)}-A_2^{(3)}\big)
\Big]\,,\\
\tilde V_1(q^2)=&\frac{N_c}{16\pi^3}\int dxd^2k_\perp\frac{h'h''}
{\bar{x}\hat{N}_1^\prime\hat{N}_1^{\prime\prime}}(-1)\Big\{\big[-16A_1^{(3)}-
2{f}(x,k_\perp,q_\perp)-4x({m'_1}+{m''_1}){m_2}\big]\nonumber\\
+&\frac{4}{D_{V'}}\big[{m'_1}(4A_1^{(3)}-A_1^{(2)})+{m''_1}A_1^{(2)}+4{m_2}A_1^{(3)}\big]
+\frac{4}{D_{V''}}\big[{m'_1}A_1^{(2)}+{m''_1}(4A_1^{(3)}-A_1^{(2)})\nonumber\\
+&4{m_2}A_1^{(3)}\big]+\frac{8}{D_{V'}D_{V''}}{f}(x,k_\perp,q_\perp)A_1^{(2)}\Big\}\,,\\
\tilde V_2(q^2)=&\frac{N_c}{16\pi^3}\int dxd^2k_\perp\frac{h'h''}
{\bar{x}\hat{N}_1^\prime\hat{N}_1^{\prime\prime}}\Big\{-16A_2^{(3)}+8A_1^{(2)}
-{m'_1}^2+{m''_1}^2-2m_2^2-q^2+2{M'}^2-2Z_2-\hat{N}_1^\prime\nonumber\\
+&\hat{N}_1^{\prime\prime}
-2{m'_1}{m''_1}+4{m'_1}{m_2}+4A_2^{(1)}\big(
m_2^2-\frac{{M'}^2+{M''}^2}{2}+\frac{1}{2}q^2+{m'_1}{m''_1}-{m'_1}{m_2}-{m''_1}{m_2}\big)\nonumber\\
+&4\big(A_2^{(1)}Z_2+\frac{{M'}^2-{M''}^2}{q^2}A_1^{(2)}\big)
+16\big(\frac{{m'_1}+{m_2}}{D_{V'}}+\frac{{m''_1}+{m_2}}{D_{V''}}\big)A_2^{(3)}
+4\big(\frac{-3{m'_1}+{m''_1}-2{m_2}}{D_{V'}}\nonumber\\
+&\frac{-{m'_1}-{m''_1}-2{m_2}}{D_{V''}}\big)A_1^{(2)}
-\big( m_2^2-\frac{{M'}^2+{M''}^2}{2}+\frac{1}{2}q^2+{m'_1}{m''_1}-{m'_1}{m_2}-{m''_1}{m_2}\big)\frac{16}{D_{V'}D_{V''}}\nonumber\\
\times&A_2^{(3)}-\frac{16}{D_{V'}D_{V''}}\big[A_2^{(3)}Z_2+\frac{{M'}^2-{M''}^2}{3q^2}(A_1^{(2)})^2\big]
+\big({m'_1}^2-{m''_1}^2+2m_2^2+q^2-2{M'}^2+\hat{N}_1^\prime\nonumber\\
-&\hat{N}_1^{\prime\prime}+2{m'_1}{m''_1}-4{m'_1}{m_2}\big)\frac{4}{D_{V'}D_{V''}}A_1^{(2)}
+\frac{8}{D_{V'}D_{V''}}A_1^{(2)}Z_2\Big\}\,,\\
\tilde V_3(q^2)=&\frac{N_c}{16\pi^3}\int dxd^2k_\perp\frac{h'h''}
{\bar{x}\hat{N}_1^\prime\hat{N}_1^{\prime\prime}}(M'^2-M''^2)\Big\{
8x(A_2^{(2)}-A_4^{(2)}+A_2^{(1)}-A_1^{(1)})
+\frac{4}{D_{V'}}\Big[{m'_1}(1-2x)(A_2^{(2)}\nonumber\\
-&A_4^{(2)}+A_2^{(1)}-A_1^{(1)})
+{m''_1}(A_2^{(2)}+A_4^{(2)}-2A_3^{(2)})+{m_2}x(2A_4^{(2)}-2A_2^{(2)}
+A_1^{(1)}-A_2^{(1)})\Big]
\nonumber\\
+&\frac{4}{D_{V''}}\Big[{m'_1}(A_2^{(2)}+2A_3^{(2)}+A_4^{(2)}-2A_2^{(1)}-2A_1^{(1)}+1)
+{m''_1}(1-2x)(A_2^{(2)}-A_4^{(2)}+A_2^{(1)}-A_1^{(1)})\nonumber\\
+&2{m_2}(2A_5^{(3)}-2A_3^{(3)}+A_2^{(2)}-3A_3^{(2)}+A_1^{(1)})\Big]
\nonumber\\
-&\frac{8}{D_{V'}D_{V''}}(A_2^{(2)}-A_4^{(2)}+A_2^{(1)}-A_1^{(1)}){f}(x,k_\perp,q_\perp)
\Big\}\,,\\
\tilde V_4(q^2)=&\frac{N_c}{16\pi^3}\int dxd^2k_\perp\frac{h'h''}
{\bar{x}\hat{N}_1^\prime\hat{N}_1^{\prime\prime}}(M''^2-M'^2)\Big\{
8(2A_4^{(3)}-2A_6^{(3)}-2A_3^{(2)}+3A_4^{(2)}+A_1^{(1)}-A_2^{(1)}-A_2^{(2)})\nonumber\\
+&4\frac{{m'_1}}{D_{V'}}(-4A_4^{(3)}+4A_6^{(3)}+4A_3^{(2)}+
3A_2^{(2)}-7A_4^{(2)}-3A_1^{(1)}+3A_2^{(1)})
+4\frac{{m''_1}}{D_{V'}}(2A_3^{(2)}-A_4^{(2)}-A_2^{(2)})\nonumber\\
+&8\frac{{m_2}}{D_{V'}}(-2A_4^{(3)}+2A_6^{(3)}+A_3^{(2)}+A_2^{(2)}
-2A_4^{(2)})
+4\frac{{m'_1}}{D_{V''}}(1-2A_1^{(1)}-2A_2^{(1)}+A_2^{(2)}+2A_3^{(2)}+A_4^{(2)})\nonumber\\
+&4\frac{{m''_1}}{D_{V''}}(-A_1^{(1)}+2A_2^{(1)}+A_2^{(2)}+4A_3^{(2)}+4A_6^{(3)}-4A_4^{(3)}-5A_4^{(2)})
+8\frac{{m_2}}{D_{V''}}(-A_1^{(1)}+2A_2^{(1)}+A_2^{(2)}\nonumber\\
+&A_3^{(2)}+2A_6^{(3)}
-2A_4^{(3)}-4A_4^{(2)})
+\frac{16}{D_{V'}D_{V''}}(-A_3^{(2)}+A_4^{(2)}+A_4^{(3)}-A_6^{(3)})\Big(m_2^2-\frac{{M'}^2+{M''}^2}{2}
\nonumber\\
+&\frac{1}{2}q^2+{m'_1}{m''_1}+{m'_1}{m_2}+{m''_1}{m_2}\Big)-\frac{8}{D_{V'}D_{V''}}(A_2^{(1)}-3A_4^{(2)}+2A_6^{(3)})Z_2
\nonumber\\
-&\frac{4}{D_{V'}D_{V''}}(A_1^{(1)}-A_2^{(1)}-A_2^{(2)}+A_4^{(2)})\big[2M'^2+(m_1'-m_1'')^2-2(m_1'+m_2)^2
-q^2-\hat{N}_1'+\hat{N}_1''\big]
\nonumber\\
-&\frac{8}{D_{V'}D_{V''}}\big[A_1^{(2)}-6A_1^{(2)}A_2^{(1)}+6A_2^{(1)}A_2^{(3)}
-2\frac{(A_1^{(2)})^2}{q^2}\big]\frac{M'^2-M''^2}{q^2}
\Big\}\,,\\
\tilde V_5(q^2)=&\frac{N_c}{16\pi^3}\int dxd^2k_\perp\frac{h'h''}
{\bar{x}\hat{N}_1^\prime\hat{N}_1^{\prime\prime}}(-1)\bigg\{
16(A_1^{(3)}-A_2^{(3)})+2(\hat{N}_1^\prime+{m'_1}^2-{M'}^2+Z_2+m_2^2-2{m'_1}{m_2})
\nonumber\\
+&4(A_2^{(1)}-A_1^{(1)})\Big(\hat{N}_1^{\prime\prime}+{m''_1}^2+\frac{{M'}^2-{M''}^2-q^2}{2}-{m'_1}{m''_1}+{m'_1}{m_2}-{m''_1}{m_2}\Big)
+\frac{4}{D_{V'}}\nonumber\\
\times&\Big[{m'_1}\Big(4(A_2^{(3)}-A_1^{(3)})+(A_2^{(1)}-A_1^{(1)})({M''}^2
-\hat{N}_1^{\prime\prime}-{m''_1}^2-m_2^2)-(A_2^{(1)}Z_2+\frac{{M'}^2-{M''}^2}{q^2}A_1^{(2)})\Big)
\nonumber\\
+&{m''_1}\Big((A_2^{(1)}-A_1^{(1)})(
\hat{N}_1^{\prime}-{M'}^2+{m'_1}^2-m_2^2)+(A_2^{(1)}Z_2+\frac{{M'}^2-{M''}^2}{q^2}A_1^{(2)})\Big)
\nonumber\\+&{m_2}\Big( 4(A_2^{(3)}-A_1^{(3)})
+(A_2^{(1)}-A_1^{(1)})(
-\hat{N}_1^{\prime}-\hat{N}_1^{\prime\prime}-{m'_1}^2-{m''_1}^2-q_\perp^2+2{m'_1}{m''_1})\Big) \Big]
\nonumber\\
+&\frac{4}{D_{V''}}\Big[2{m'_1}A_1^{(2)}+4{m''_1}4(A_2^{(3)}-A_1^{(3)})+{m_2}\Big(4(A_2^{(3)}-A_1^{(3)})-2A_1^{(2)}\Big)\Big]
\nonumber\\
+&\frac{16}{D_{V'}D_{V''}}\Big[(A_2^{(3)}-A_1^{(3)})
\big(-{m'_1}{m''_1}\!-m_2^2\!-{m'_1}{m_2}\!-{m''_1}{m_2}-\frac{{M'}^2\!-{M''}^2\!+
q^2}{2}\!+{M'}^2\big)\nonumber\\
-&A_2^{(3)}Z_2-\frac{{M'}^2-{M''}^2}{3q^2}(A_1^{(2)})^2\Big]
\!\bigg\}\,,\\
\tilde V_6(q^2)=&\frac{N_c}{16\pi^3}\int dxd^2k_\perp\frac{h'h''}
{\bar{x}\hat{N}_1^\prime\hat{N}_1^{\prime\prime}}\bigg\{
16(A_1^{(2)}-A_1^{(3)}-A_2^{(3)})+2(
-2{m'_1}^2-{m''_1}^2-m_2^2+q^2+{M'}^2-Z_2\nonumber\\
-&2\hat{N}_1^\prime-\hat{N}_1^{\prime\prime}
+2{m'_1}{m''_1}+2{m'_1}{m_2})
+4(A_2^{(1)}+A_1^{(1)})\Big({m'_1}^2+\frac{{M'}^2-{M''}^2}{2}+\hat{N}_1^\prime-\frac{1}{2}q^2-{m'_1}{m''_1}\nonumber\\
-&{m'_1}{m_2}-{m''_1}{m_2}\Big)
+\frac{16}{D_{V'}}\Big[({m'_1}+{m_2})(A_1^{(3)}+A_2^{(3)})-({m'_1}+\frac{1}{2}{m''_1}+
\frac{1}{2}{m_2})A_1^{(2)}
\Big]\nonumber\\
+&4\frac{{m'_1}}{D_{V''}}(-{m''_1}^2-m_2^2+{M''}^2-Z_2-\hat{N}_1^{\prime\prime})
+\frac{16}{D_{V''}}({m''_1}+{m_2})(A_1^{(3)}+A_2^{(3)}-A_1^{(2)})
\nonumber\\
+&\frac{4}{D_{V''}}\Big[{m''_1}({m'_1}^2+m_2^2-{M'}^2+Z_2+\hat{N}_1^{\prime})
+{m_2}({m'_1}^2+{m''_1}^2-q^2+\hat{N}_1^{\prime}+\hat{N}_1^{\prime\prime}-2{m'_1}{m''_1})\Big]\nonumber\\
+&\frac{4}{D_{V''}}(A_1^{(1)}+A_2^{(1)})\Big[{m'_1}({m''_1}^2+m_2^2-{M''}^2+\hat{N}_1^{\prime\prime})
+{m''_1}(-{m'_1}^2-m_2^2+{M'}^2-\hat{N}_1^{\prime})\nonumber\\
+&{m_2}(-{m'_1}^2-{m''_1}^2+q^2-\hat{N}_1^{\prime}-\hat{N}_1^{\prime\prime}+2{m'_1}{m''_1})\Big]
+\frac{4}{D_{V''}}({m'_1}-{m''_1})\Big(A_2^{(1)}Z_2+\frac{{M'}^2-{M''}^2}{q^2}A_1^{(2)}\Big)
\nonumber\\
-&\frac{16}{D_{V'}D_{V''}}(A_1^{(3)}+A_2^{(3)}-A_1^{(2)})
\Big(m_2^2-\frac{{M'}^2+{M''}^2}{2}+\frac{1}{2}q^2+{m'_1}{m''_1}+{m'_1}{m_2}+{m''_1}{m_2}\Big)
\nonumber\\
-&\frac{16}{D_{V'}D_{V''}}\Big[A_2^{(3)}Z_2+\frac{{M'}^2-{M''}^2}{3q^2}(A_1^{(2)})^2-A_1^{(2)}Z_2\Big]
\bigg \}\,,
\end{align}
where ${f}(x,k_\perp,q_\perp)=\frac{x^2}{\bar{x}}m_2^2+\frac{1}{\bar{x}}k_\perp^2-k_\perp\cdot q_\perp+\bar{x}m'_1m''_1-x(m'_1m_2+m''_1m_2)$ and $D_{V^{\prime(\prime\prime)}}={M}^{\prime(\prime\prime)}_0+m_1^{\prime(\prime\prime)}+m_2$ is the factor appeared in the vertex operator. Here, we use the same notation and convention as Refs.~\cite{Jaus:1999zv,Jaus:2002sv,Cheng:2003sm}, and the explicit forms of $Z_2$, $h^{\prime(\prime\prime)}/\hat{N}_1^{\prime(\prime\prime)}$ and $A_i^{(j)}$  functions can be easily found therein. 

\end{appendix}

\section*{Acknowledgements}
This work is supported by the National Natural Science Foundation of China (Grant Nos. 11875122 and 11475055) and the Program for Innovative Research Team in University of Henan Province (Grant No.19IRTSTHN018).


\begin{thebibliography}{99}

 \bibitem{Bediaga:2012py}
  R.~Aaij {\it et al.} [LHCb Collaboration],
  Eur.\ Phys.\ J.\ C {\bf 73} (2013) no.4,  2373.

  \bibitem{Abe:2010gxa}
  T.~Abe {\it et al.} [Belle-II Collaboration],
  arXiv:1011.0352 [physics.ins-det].

 \bibitem{Tanabashi:2018oca}
  M.~Tanabashi {\it et al.} [Particle Data Group],
  Phys.\ Rev.\ D {\bf 98} (2018) no.3,  030001.

  \bibitem{Isgur:1991wq}
  N.~Isgur and M.~B.~Wise,
  Phys.\ Rev.\ Lett.\  {\bf 66} (1991) 1130.

  \bibitem{Godfrey:1986wj}
  S.~Godfrey and R.~Kokoski,
  Phys.\ Rev.\ D {\bf 43} (1991) 1679.

  \bibitem{Eichten:1993ub}
  E.~J.~Eichten, C.~T.~Hill and C.~Quigg,
  Phys.\ Rev.\ Lett.\  {\bf 71} (1993) 4116.

  \bibitem{Ebert:1997nk}
  D.~Ebert, V.~O.~Galkin and R.~N.~Faustov,
  Phys.\ Rev.\ D {\bf 57} (1998) 5663
   Erratum: [Phys.\ Rev.\ D {\bf 59} (1999) 019902].
   
  \bibitem{Aaij:2014jba}
  R.~Aaij {\it et al.} [LHCb Collaboration],
  Int.\ J.\ Mod.\ Phys.\ A {\bf 30} (2015) no.07,  1530022.

  \bibitem{Chang:2015jla}
  Q.~Chang, P.~P.~Li, X.~H.~Hu and L.~Han,
  Int.\ J.\ Mod.\ Phys.\ A {\bf 30} (2015) no.27,  1550162.


  \bibitem{Aaij:2010gn}
  R.~Aaij {\it et al.} [LHCb Collaboration],
  Phys.\ Lett.\ B {\bf 694} (2010) 209.
 
   \bibitem{Grinstein:2015aua}
  B.~Grinstein and J.~Martin Camalich,
  Phys.\ Rev.\ Lett.\  {\bf 116} (2016) no.14,  141801.
     
  \bibitem{Sun:2019xyw}
  H.~K.~Sun and M.~Z.~Yang,
  Phys.\ Rev.\ D {\bf 99} (2019) no.9,  093002.
 
 \bibitem{Xu:2015eev}
  G.~Z.~Xu, Y.~Qiu, C.~P.~Shen and Y.~J.~Zhang,
  Eur.\ Phys.\ J.\ C {\bf 76} (2016) no.11,  583.

\bibitem{Saini:2019tge}
  J.~Saini, D.~Kumar, S.~Gangal and S.~B.~Das,
  arXiv:1905.03933 [hep-ph].  
  
\bibitem{Kumbhakar:2018uty}
  S.~Kumbhakar and J.~Saini,
  Eur.\ Phys.\ J.\ C {\bf 79} (2019) no.5,  394.

\bibitem{Banerjee:2017upm}
  D.~Banerjee, P.~Maji and S.~Sahoo,
  Int.\ J.\ Mod.\ Phys.\ A {\bf 32} (2017) no.14,  1750075.

\bibitem{Sahoo:2017zke}
  S.~Sahoo and R.~Mohanta,
  Springer Proc.\ Phys.\  {\bf 203} (2018) 321.

  
   \bibitem{Wang:2012hu}
  Z.~G.~Wang,
  Commun.\ Theor.\ Phys.\  {\bf 61} (2014) no.1,  81.

  \bibitem{Zeynali:2014wya}
  K.~Zeynali, V.~Bashiry and F.~Zolfagharpour,
  Eur.\ Phys.\ J.\ A {\bf 50} (2014) 127.

  \bibitem{Bashiry:2014qia}
  V.~Bashiry,
  Adv.\ High Energy Phys.\  {\bf 2014} (2014) 503049.

\bibitem{Chang:2016cdi}
  Q.~Chang, J.~Zhu, X.~L.~Wang, J.~F.~Sun and Y.~L.~Yang,
  Nucl.\ Phys.\ B {\bf 909} (2016) 921.

\bibitem{Wang:2018dtb}
  T.~Wang, Y.~Jiang, T.~Zhou, X.~Z.~Tan and G.~L.~Wang,
  J.\ Phys.\ G {\bf 45} (2018) no.11,  115001.
        
   \bibitem{Chang:2018sud}
  Q.~Chang, J.~Zhu, N.~Wang and R.~M.~Wang,
  Adv.\ High Energy Phys.\  {\bf 2018} (2018) 7231354.  


\bibitem{Chang:2015ead}
  Q.~Chang, X.~Hu, J.~Sun, X.~Wang and Y.~Yang,
  Adv.\ High Energy Phys.\  {\bf 2015} (2015) 767523.
  
  \bibitem{Sun:2017xed}
  J.~Sun, H.~Li, Y.~Yang, N.~Wang, Q.~Chang and G.~Lu,
  J.\ Phys.\ G {\bf 44} (2017) no.7,  075007.
  
  \bibitem{Chang:2016meh}
  Q.~Chang, X.~Hu, Z.~Chang, J.~Sun and Y.~Yang,
  Adv.\ High Energy Phys.\  {\bf 2016} (2016) 3863725.
  doi:10.1155/2016/3863725
    
  \bibitem{Chang:2016eto}
  Q.~Chang, L.~X.~Chen, Y.~Y.~Zhang, J.~F.~Sun and Y.~L.~Yang,
  Eur.\ Phys.\ J.\ C {\bf 76} (2016) no.10,  523.

  \bibitem{Buchalla:1995vs}
  G.~Buchalla, A.~J.~Buras and M.~E.~Lautenbacher,
  Rev.\ Mod.\ Phys.\  {\bf 68} (1996) 1125.

\bibitem{Buras:1998raa}
  A.~J.~Buras,
  hep-ph/9806471.


  \bibitem{Fakirov:1977ta}
  D.~Fakirov and B.~Stech,
  Nucl.\ Phys.\ B {\bf 133} (1978) 315.

 \bibitem{Cabibbo:1977zv}
  N.~Cabibbo and L.~Maiani,
  Phys.\ Lett.\  {\bf 73B} (1978) 418
   Erratum: [Phys.\ Lett.\  {\bf 76B} (1978) 663].

\bibitem{Bauer:1984zv}
  M.~Bauer and B.~Stech,  Phys.\ Lett.\ B {\bf 152} (1985) 380.

\bibitem{Wirbel:1985ji}
  M.~Wirbel, B.~Stech and M.~Bauer, Z.\ Phys.\ C {\bf 29} (1985) 637.

\bibitem{Bauer:1986bm}
  M.~Bauer, B.~Stech and M.~Wirbel,  Z.\ Phys.\ C {\bf 34} (1987) 103.

 \bibitem{Bjorken:1988kk}
  J.~D.~Bjorken,  Nucl.\ Phys.\ Proc.\ Suppl.\  {\bf 11} (1989) 325.

  \bibitem{Jain:1995dd}
  P.~Jain, B.~Pire and J.~P.~Ralston, Phys.\ Rept.\  {\bf 271} (1996) 67.


 \bibitem{Beneke:1999br}
  M.~Beneke, G.~Buchalla, M.~Neubert and C.~T.~Sachrajda,
  Phys.\ Rev.\ Lett.\  {\bf 83} (1999) 1914.

  \bibitem{Beneke:2000ry}
  M.~Beneke, G.~Buchalla, M.~Neubert and C.~T.~Sachrajda,
  Nucl.\ Phys.\ B {\bf 591} (2000) 313.


  

\bibitem{Huber:2016xod}
  T.~Huber, S.~Krankl and X.~Q.~Li,
  JHEP {\bf 1609} (2016) 112.

 \bibitem{Wang:2007ys}
  Y.~M.~Wang, H.~Zou, Z.~T.~Wei, X.~Q.~Li and C.~D.~Lu,
  Eur.\ Phys.\ J.\ C {\bf 54} (2008) 107.



 \bibitem{Lubicz:2016bbi}
  V.~Lubicz, A.~Melis and S.~Simula,
  PoS LATTICE {\bf 2016} (2017) 291.

 \bibitem{Straub:2015ica}
  A.~Bharucha, D.~M.~Straub and R.~Zwicky,
  JHEP {\bf 1608} (2016) 098.

 \bibitem{Braun:2016wnx}
  V.~M.~Braun {\it et al.},
  JHEP {\bf 1704} (2017) 082.



  \bibitem{Goity:2000dk}
  J.~L.~Goity and W.~Roberts,
  Phys.\ Rev.\ D {\bf 64} (2001) 094007.

  \bibitem{Ebert:2002xz}
  D.~Ebert, R.~N.~Faustov and V.~O.~Galkin,
  Phys.\ Lett.\ B {\bf 537} (2002) 241.

  \bibitem{Zhu:1996qy}
  S.~L.~Zhu, W.~Y.~P.~Hwang and Z.~s.~Yang,
  Mod.\ Phys.\ Lett.\ A {\bf 12} (1997) 3027.


  \bibitem{Aliev:1995zlh}
  T.~M.~Aliev, D.~A.~Demir, E.~Iltan and N.~K.~Pak,
  Phys.\ Rev.\ D {\bf 54} (1996) 857.

  \bibitem{Colangelo:1993zq}
  P.~Colangelo, F.~De Fazio and G.~Nardulli,
  Phys.\ Lett.\ B {\bf 316} (1993) 555.

  \bibitem{Choi:2007se}
  H.~M.~Choi,
  Phys.\ Rev.\ D {\bf 75} (2007) 073016.

  \bibitem{Cheung:2014cka}
  C.~Y.~Cheung and C.~W.~Hwang,
  JHEP {\bf 1404} (2014) 177.

 \bibitem{Chang:2018zjq}
  Q.~Chang, X.~N.~Li, X.~Q.~Li, F.~Su and Y.~D.~Yang,
  Phys.\ Rev.\ D {\bf 98} (2018) no.11,  114018.

  \bibitem{Chang:2019mmh}
  Q.~Chang, X.~N.~Li and L.~T.~Wang,
  Eur.\ Phys.\ J.\ C {\bf 79} (2019) no.5,  422.

 \bibitem{Jaus:1999zv}
  W.~Jaus,
  Phys.\ Rev.\ D {\bf 60}  (1999) 054026.

\bibitem{Jaus:2002sv}
  W.~Jaus,
  Phys.\ Rev.\ D {\bf 67}  (2003)  094010.

\bibitem{Cheng:2003sm}
  H.~Y.~Cheng, C.~K.~Chua and C.~W.~Hwang,
  Phys.\ Rev.\ D {\bf 69} (2004) 074025.

 \bibitem{Ali:1978kn}
  A.~Ali, J.~G.~Korner, G.~Kramer and J.~Willrodt,
  Z.\ Phys.\ C {\bf 1} (1979) 269.

  \bibitem{Korner:1979ci}
  J.~G.~Korner and G.~R.~Goldstein,
  Phys.\ Lett.\ B {\bf 89} (1979) 105.

 \bibitem{Kagan:2004uw}
  A. L. Kagan, Phys. Lett. B {\bf 601} (2004) 151.

\bibitem{Jaus:1996np}
  W.~Jaus,
  Phys.\ Rev.\ D {\bf 53} (1996) 1349
  Erratum: [Phys.\ Rev.\ D {\bf 54} (1996) 5904].  
 
 \bibitem{Barik:1997qq}
  N.~Barik, S.~K.~Tripathy, S.~Kar and P.~C.~Dash,
  Phys.\ Rev.\ D {\bf 56} (1997) 4238.
   
\bibitem{Chang:2017sdl}
  Q.~Chang, S.~Xu and L.~Chen,
  Nucl.\ Phys.\ B {\bf 921} (2017) 454.
  
  \bibitem{Wang:2017mqp}
  W.~Wang, F.~S.~Yu and Z.~X.~Zhao,
  Eur.\ Phys.\ J.\ C {\bf 77} (2017) no.11,  781.
  
 \bibitem{Choi:2009ai}
  H.~M.~Choi and C.~R.~Ji,
  Phys.\ Rev.\ D {\bf 80} (2009) 054016.  

 \bibitem{Choi:2009ym}
  H.~M.~Choi and C.~R.~Ji,
  Phys.\ Rev.\ D {\bf 80} (2009) 114003.
    
\bibitem{Choi:2010ha}
  H.~M.~Choi,
  Phys.\ Rev.\ D {\bf 81} (2010) 054003.
  
\bibitem{Choi:2010zb}
  H.~M.~Choi,
  J.\ Phys.\ G {\bf 37} (2010) 085005.

  \bibitem{Hwang:2001wu}
  C.~W.~Hwang,
  Phys.\ Lett.\ B {\bf 530} (2002) 93.
  
\bibitem{Wang:2008ci}
  W.~Wang and Y.~L.~Shen,
  Phys.\ Rev.\ D {\bf 78} (2008) 054002.  

\bibitem{Wang:2008xt}
  W.~Wang, Y.~L.~Shen and C.~D.~Lu,
  Phys.\ Rev.\ D {\bf 79} (2009) 054012.

\bibitem{Shen:2008zzb}
  Y.~L.~Shen and Y.~M.~Wang,
  Phys.\ Rev.\ D {\bf 78} (2008) 074012.
  
  \bibitem{Wang:2009mi}
  X.~X.~Wang, W.~Wang and C.~D.~Lu,
  Phys.\ Rev.\ D {\bf 79} (2009) 114018.
 
 \bibitem{Cheng:2017pcq}
  H.~Y.~Cheng and X.~W.~Kang,
  Eur.\ Phys.\ J.\ C {\bf 77} (2017) no.9,  587
   Erratum: [Eur.\ Phys.\ J.\ C {\bf 77} (2017) no.12,  863].

\bibitem{Kang:2018jzg}
  X.~W.~Kang, T.~Luo, Y.~Zhang, L.~Y.~Dai and C.~Wang,
  Eur.\ Phys.\ J.\ C {\bf 78} (2018) no.11,  909.

       
\end{thebibliography}
\end{document}